\documentclass[aps,prb,twocolumn
,showpacs]{revtex4}

\usepackage{amsmath,amssymb}
\usepackage{graphicx}
\usepackage{dcolumn}
\usepackage{bm}

\def\rnum#1{\expandafter{%
\romannumeral #1}}
\def\Rnum#1{\uppercase\expandafter{%
\romannumeral #1}}

\begin{document}


\title{Four-spin-exchange- and magnetic-field-induced chiral order in
two-leg spin ladders}

\author{Masahiro Sato}
\affiliation{Synchrotron Radiation Research Center, 
Japan Atomic Energy Agency, Sayo, Hyogo 679-5148, Japan\\
CREST Japan Science and
Technology Agency, Kawaguchi, Saitama 332-0012, Japan\\
Condensed Matter Theory Laboratory, RIKEN, Wako, Saitama 351-0198, Japan}



\date{\today}

\begin{abstract}
We propose a mechanism of a vector chiral long-range order 
in two-leg spin-$\frac{1}{2}$ and spin-1 antiferromagnetic 
ladders with four-spin exchanges and a Zeeman term. 
It is known that for one-dimensional quantum systems, 
spontaneous breakdown of continuous symmetries is generally 
forbidden. Any vector chiral order hence does not appear in 
spin-rotationally [SU(2)]-symmetric spin ladders. However, if a 
magnetic field is added along the $S^z$ axis of ladders 
and the SU(2) symmetry is reduced to the U(1) one, the $z$ component of 
a vector chiral order can emerge 
with the remaining U(1) symmetry unbroken. 
Making use of Abelian bosonization techniques, we actually 
show that a certain type of four-spin exchange can 
yield a vector chiral long-range order in spin-$\frac{1}{2}$ and spin-1 
ladders under a magnetic field. 
In the chiral-ordered phase, the $Z_2$ interchain-parity 
(i.e., chain-exchange) symmetry is spontaneously broken. We also 
consider effects of perturbations breaking the parity symmetry.   
\end{abstract}


\maketitle

\section{Introduction} 
Quantum spin ladder systems have attracted much 
attention for more than a decade.~\cite{Gogo} There are numerous
compounds with a ladder structure. 
It is well known now that ladders contain considerably 
rich physics: for example, 
they provide fascinating ordered states,~\cite{Lu-Tr,Hi-Mo,Le-To,N_T-K_M} 
string order parameters detecting different topological 
sectors,~\cite{Nishi-Naka} interesting exactly solvable 
points,~\cite{M-G,Aza} etc. 
In particular, physics of two-leg ladders has been greatly developed. 
Low-energy properties of nonfrustrated two-leg ladders have been
well understood: as soon as an infinitesimal antiferromagnetic (AF)
[ferromagnetic (FM)] interchain exchange is added between 
two spin-$\frac{1}{2}$ AF critical chains, 
a gap opens and a rung-singlet [Haldane] type 
spin-liquid state occurs,~\cite{Nishi-Naka,Shel} 
while even if an interchain coupling with an arbitrary strength 
is introduced between two spin-1 AF Haldane-gapped chains, 
the gap always survives.~\cite{S1ladder} 
Based on these properties of standard ladders, 
many physicists recently have been focusing on more realistic or 
frustrated ladder systems. 
Among them, effects of four-spin exchanges 
have been studied intensively now.  
Although it is generally difficult to understand effects of 
four-spin exchanges compared with conventional two-spin ones, 
thanks to the 
discovery of a duality relation in two-leg spin-$\frac{1}{2}$ 
ladders,~\cite{Hi-Mo} powerful analytical tools 
(conformal field theory, Bethe ansatz, etc.) 
and accurate numerical methods in (1+1) dimensions, 
physical properties of four-spin exchanges in ladders have been 
gradually elucidated.~\cite{M-K} For instance, it has been shown that 
in spin-rotationally [SU(2)]-symmetric spin-$\frac{1}{2}$ ladders, 
four-spin exchanges can cause novel phases: a staggered dimer order, 
a scalar chiral one, and a spin liquid with a strong (but short-range) 
vector chirality correlation.~\cite{Lu-Tr,Hi-Mo,Le-To} Moreover, 
in SU(2)-symmetric spin-1 ladders with a biquadratic term, 
a certain parameter region where a spin-nematic correlation 
is dominant is predicted.~\cite{Lu-Sch}

In this paper, motivated by the above studies of spin ladders, 
we investigate two-leg spin-$\frac{1}{2}$ and spin-1 
AF ladders with four-spin exchanges 
{\it in the presence of a magnetic field}, 
which reduces the SU(2) symmetry to a U(1) one. 
For these ladders, we propose a mechanism of a vector chiral 
long-range order which spontaneously breaks the parity symmetry 
for the interchain (rung) direction. 

The paper is organized as follows. 
Throughout the paper, we apply the effective field-theory approach 
and focus on the weak two- and four-spin rung-coupling regime.  
In Sec.~\ref{bos_review}, 
we first review simple spin-$\frac{1}{2}$ ladders without
four-spin exchanges under a magnetic field using Abelian bosonization 
techniques and discuss the physical meaning of locking the bosonic
fields. Then, based on the discussion, 
we prove the mechanism of a vector chiral order in the next 
two sections: Secs.~\ref{4spin_spin1_2} and \ref{4spin_spin1} are 
devoted to investigating spin-$\frac{1}{2}$ and spin-1 cases, 
respectively. In Sec.~\ref{chiral_break}, we shortly consider 
a few perturbation terms breaking the rung-parity symmetry: 
celebrated Dzyaloshinsky-Moriya (DM) interactions~\cite{DMterm} 
and a parity-breaking three-spin term.~\cite{three_spin,three_spin_2}  
Finally, we summarize our results and then discuss them in 
Sec.~\ref{final}.

\section{Bosonization for spin-$\frac{\bm 1}{\bm 2}$ ladders}
\label{bos_review}
We begin with a fundamental 
spin-$\frac{1}{2}$ AF ladder,~\cite{Gogo,M-K} whose Hamiltonian is
\begin{eqnarray}
\label{S=1/2ladder}  
{\cal H}_{\rm lad}&=&J
\sum_{n,j}\bm{S}_{n,j}\cdot\bm{S}_{n,j+1} 
+J_\perp\sum_{j}\bm{S}_{1,j}\cdot\bm{S}_{2,j}\nonumber\\
&&
-H\sum_{n,j}S_{n,j}^z, 
\end{eqnarray}
where $\bm{S}_{n,j}$ is the spin-$\frac{1}{2}$ operator on site $j$ of 
the $n$th chain ($n=1,2$), 
$J>0$ ($J_\perp$) is the intrachain (rung) coupling, 
and $H$ is a magnetic field. 
At the decoupled point of $J_\perp=0$, 
the low-energy effective Hamiltonian for the $n$th chain is 
a Gaussian model 
\begin{eqnarray*}
{\cal H}_n^{\rm eff}=\int dx\frac{v}{2}
[K^{-1}(\partial_x\phi_n)^2+K(\partial_x\theta_n)^2],
\end{eqnarray*} 
where $x=ja_0$ ($a_0$ is the lattice constant), $\phi_n$ is the scalar boson 
field, and $\theta_n$ is the dual to $\phi_n$: these two fields 
satisfy a canonical commutation relation 
$[\phi_n(x),\partial_y\theta_n(y)]=i\delta(x-y)$. 
Quantities $v$ and $K$
denote the spin-wave velocity and the Tomonaga-Luttinger liquid (TLL) 
parameter, respectively. The value of $K$ ($v$) runs from 
$1/2$ ($\pi Ja_0/2$) to 1 (0)
with $H$ increasing from $0$ to the upper critical field $2J$. 
Spin operators are also
bosonized as 
\begin{subequations}
\label{boso_spin_1/2} 
\begin{eqnarray}
S^z_{n,j}&\approx & M+\frac{a_0}{\sqrt{\pi}}\partial_x{\phi_n}
\nonumber\\
&&+(-1)^j A_1 \sin(\sqrt{4\pi}{\phi_n}+2\pi Mj)+\cdots,
\\
S^+_{n,j}&\approx& e^{i\sqrt{\pi}\theta_n}[(-1)^j B_0
\nonumber\\
&&+B_1\sin(\sqrt{4\pi}{\phi_n}+2\pi M j)+\cdots],
\end{eqnarray}
\end{subequations} 
where $M(H)=\langle S_{n,j}^z \rangle$, 
and $A_q$ and $B_q$ are nonuniversal constants.~\cite{Lu-Hi} 
From this bosonization framework, 
the low-energy theory of the ladder~(\ref{S=1/2ladder}) with
$|J_\perp|\ll J$ is written as 
\begin{eqnarray}
\label{S=1/2eff}  
{\cal H}_{\rm eff}&=& \int dx \sum_{t=\pm}\frac{v_t}{2}
\left[K_t^{-1}(\partial_x\phi_t)^2+K_t(\partial_x\theta_t)^2\right]
\nonumber\\
&&+\frac{J_\perp}{a_0}\Big[B_0^2\cos(\sqrt{2\pi}\theta_-)+
\frac{A_1^2}{2}\cos(\sqrt{8\pi}\phi_-)
\nonumber\\
&&-\frac{A_1^2}{2}\cos(\sqrt{8\pi}\phi_++4\pi Mj)\Big]
+\sqrt{\frac{2}{\pi}}MJ_\perp \partial_x \phi_+
\nonumber\\ 
&&+\cdots,
\end{eqnarray}
where we introduced new fields 
$\phi_\pm=(\phi_1\pm\phi_2)/\sqrt{2}$ and 
$\theta_\pm=(\theta_1\pm\theta_2)/\sqrt{2}$ that obey 
$[\phi_\pm(x),\partial_y\theta_\pm(y)]=i\delta(x-y)$. 
New TLL parameters $K_\pm$ and velocities $v_\pm$ are
estimated as 
\begin{eqnarray}
\label{TLLpara_velocity}
K_\pm & \approx & K\Big(1\mp K\frac{J_\perp a_0}{2\pi v}\Big), \nonumber\\
v_\pm & \approx & v\Big(1\pm K\frac{J_\perp a_0}{2\pi v}\Big).
\end{eqnarray} 
The boson linear term $\partial_x\phi_+$ can be absorbed into 
the Gaussian (boson quadratic) part by the shift 
$\tilde \phi_+ =\phi_+ + \sqrt{\frac{2}{\pi}}K_+ M 
\frac{J_\perp x}{v_+}$ which accompanies the correction of the
magnetization, $\langle S_{n,j}\rangle =
(1-\frac{J_\perp a_0}{\pi v_+}K_+)M=\tilde M$. 
Hereafter, we will again use symbols $\phi_+$ and $M$ for $\tilde\phi_+$
and $\tilde M$, respectively.

The U(1) spin rotation around the $S^z$ axis, 
the translation by one site for the chain direction, the site-parity
operation for the same direction and the rung-parity
transformation (i.e., exchange between two chains) correspond to 
$\theta_+\to\theta_+ +{\rm const}$, 
$(\phi_+,\theta_+)\to(\phi_+ +\sqrt{2\pi}(M+1/2),
\theta_+ +\sqrt{2\pi})$, 
$\big(\phi_+(x),\phi_-(x),\theta_\pm(x)\big)\to
\big(-\phi_+(-x)+\sqrt{\pi/2},-\phi_-(-x),\theta_\pm(-x)\big)$ 
and $(\phi_-,\theta_-)\to-(\phi_-,\theta_-)$, respectively. 
The Hamiltonian~(\ref{S=1/2eff})
hence does not contain $\theta_+$ vertex terms, $\phi_+$ ones
without spatially oscillating factors $e^{\pm in4\pi Mj}$, 
and sine terms with $\phi_\pm$ and $\theta_-$.

The form of Eq.~(\ref{S=1/2eff}) tells us that the low-energy physics 
of the ladder~(\ref{S=1/2ladder}) can be described by two parts, 
$(\phi_+,\theta_+)$ and $(\phi_-,\theta_-)$ sectors. 
In the present notation, scaling dimensions of 
$e^{in\sqrt{2\pi}\theta_t}$ and $e^{im\sqrt{8\pi}\phi_t}$
are, respectively, $n^2/(2 K_t)$ and $2 m^2 K_t$. 
The $(\phi_-,\theta_-)$ sector always takes a massive
spectrum due to relevant cosine terms, irrespective of the value of $M$.  
Since $\cos(\sqrt{2\pi}\theta_-)$ is usually more relevant than 
$\cos(\sqrt{8\pi}\phi_-)$ in the ladder, namely $K_->1/2$,~\cite{Sato} 
the dual field $\theta_-$ is pinned. For the AF-rung case, 
$\langle\theta_-\rangle=\pm\sqrt{\pi/2}$, whereas for
the FM-rung case, $\langle\theta_-\rangle=0$.
On the other hand, physics of the $(\phi_+,\theta_+)$ sector 
depends on $M$. When $M=0$, i.e., $H$ is smaller than the lower 
critical field $H_l$ of the ladder,~\cite{LowerCritical} 
$\cos(\sqrt{8\pi}\phi_+)$ is relevant and a spin gap thus emerges. 
This state is nothing but the Haldane or rung-singlet spin liquids. 
It is known that pinning the field $\phi_+$ corresponds to the 
emergence of nonlocal string orders.~\cite{Nishi-Naka}
Inversely, if $M\neq0$ ($H>H_l$), $\cos(\sqrt{8\pi}\phi_+ +4\pi Mj)$ is 
irrelevant due to the factor $e^{\pm i4\pi Mj}$ and the gap vanishes. 
Therefore in the case of $M\neq0$, the $(\phi_+,\theta_+)$ sector 
is described by a Gaussian model, which indeed corresponds to
the field-induced TLL phase. 
These scenarios in the ladder~(\ref{S=1/2ladder}) would be robust
against small perturbations conserving symmetries of the ladder 
(e.g., $XXZ$ anisotropy, further-neighbor exchanges, etc.).

As typical order parameters for two-leg ladder systems, 
one can find three quantities: 
the $z$ component of spin vector chirality 
${\cal V}_j=({\bm S}_{1,j}\times{\bm S}_{2,j})^z$ 
and two magnetic moments 
${\cal N}_{\pm,j}=S_{1,j}^z\pm S_{2,j}^z$. It is important to note that 
${\cal V}_j$ and ${\cal N}_{-,j}$ are odd for the rung-parity operation, 
while ${\cal N}_{+,j}$ is
even. In the SU(2)-symmetric case with $H=0$, which corresponds to the
gapped spin-liquid state, expectation values of these order parameters 
must be zero because the presence of them means the spontaneous
breakdown of the continuous SU(2) symmetry and such a symmetry breaking
is generally forbidden by Mermin-Wagner-Hohenberg argument.~\cite{M-W-H} 
The gapped spin-liquid state still remains if $H<H_l$ and $M=0$. 
Meanwhile, in the U(1)-symmetric case with $M\neq0$ 
their finite expectation values can be allowed 
since they do not violate the U(1) symmetry. 
Formula~(\ref{boso_spin_1/2}) enables us to 
represent these order parameters as 
\begin{subequations}
\label{orderpara_boso}
\begin{eqnarray}
{\cal V}_j &\approx&-B_0^2\sin(\sqrt{2\pi}\theta_-)+\cdots,
\label{orderpara_boso_1}\\
{\cal N}_{-,j} &\approx&(-1)^j2A_1\cos(\sqrt{2\pi}\phi_+ +2\pi Mj)
\sin(\sqrt{2\pi}\phi_-)\nonumber\\
&&+a_0\sqrt{\frac{2}{\pi}}\partial_x\phi_-+\cdots,
\label{orderpara_boso_2}\\
{\cal N}_{+,j} &\approx&(-1)^j2A_1\sin(\sqrt{2\pi}\phi_+ +2\pi Mj)
\cos(\sqrt{2\pi}\phi_-)\nonumber\\
&&+2M+a_0\sqrt{\frac{2}{\pi}}\partial_x\phi_++\cdots.
\label{orderpara_boso_3}
\end{eqnarray}
\end{subequations}
We see that ${\cal V}_j$ and ${\cal N}_{-,j}$ are actually odd for the
rung-parity transformation $(\phi_-,\theta_-)\to -(\phi_-,\theta_-)$,
whereas ${\cal N}_{+,j}$ is even. 
The leading part of ${\cal N}_{\pm}$ consists of two fields $\phi_\pm$. 
As we mentioned above, since $\phi_+$ is not locked 
in the case of $M\neq 0$, a magnetic order with 
$\langle{\cal N}_{\pm}\rangle\neq0$ is shown to disappear in
the case. On the other hand, remarkably, the leading term of 
${\cal V}_j$ contains only the field $\theta_-$ that 
can be locked even in the U(1)-symmetric case with $M\neq 0$.

From the above discussion on order parameters, 
it is inferred that the vector chiral 
order emerges in a certain class of U(1)-symmetric ladders under 
a magnetic field. For the case of the simple ladder~(\ref{S=1/2ladder}), 
we find $\langle {\cal V}_j\rangle=0$ because of 
$\langle\theta_-\rangle=0$ or $\pm\sqrt{\pi/2}$. 
However, in other words, it is expected that if the $\theta_-$-locked position
is moved even slightly by additional perturbations, the chiral order would
appear. We show below that some types of four-spin exchanges are indeed 
such a desirable perturbation.

Here, we again remark on Eq.~(\ref{orderpara_boso}). As one saw above,
for the simple spin-$\frac{1}{2}$ ladder~(\ref{S=1/2ladder}) with $M=0$, 
boson fields in both $(\phi_+,\theta_+)$ and $(\phi_-,\theta_-)$
sectors are locked. The physical meanings of locking $\phi_-$ or
$\theta_-$, however, have not been discussed well so far. 
Equation~(\ref{orderpara_boso}) clearly shows that locking $\theta_-$
can induce the vector chirality, while locking $\phi_-$ can do 
magnetic orders. Moreover, one finds that Eq.~(\ref{orderpara_boso_1})
is analogous to the celebrated supercurrent formula in 
a Josephson junction.~\cite{Nishiyama}

\section{Four-spin exchanges in U(1)-symmetric 
spin-$\frac{\bm 1}{\bm 2}$ ladders}
\label{4spin_spin1_2} 
As additional four-spin exchanges to the U(1)-symmetric spin 
ladder~(\ref{S=1/2ladder}), 
let us consider the following terms:
\begin{subequations}
\label{4spin_terms}
\begin{eqnarray} 
{\cal H}_{\rm ll}=V_{\rm ll}
\sum_j(\bm{S}_{1,j}\cdot\bm{S}_{1,j+1})(\bm{S}_{2,j}\cdot\bm{S}_{2,j+1}),
\\
{\cal H}_{\rm rr}=V_{\rm rr}
\sum_j(\bm{S}_{1,j}\cdot\bm{S}_{2,j})(\bm{S}_{1,j+1}\cdot\bm{S}_{2,j+1}),
\\
{\cal H}_{\times}=V_{\times}
\sum_j(\bm{S}_{1,j}\cdot\bm{S}_{2,j+1})(\bm{S}_{2,j}\cdot\bm{S}_{1,j+1}).
\end{eqnarray}
\end{subequations} 
The well known four-spin cyclic term satisfies
$V_{\rm ll}=V_{\rm rr}=-V_\times$.~\cite{Hi-Mo} 
The bosonized spin operators~(\ref{boso_spin_1/2}) 
and symmetry arguments provide the
following bosonized expressions of four-spin exchanges:
\begin{subequations}
\label{4spin_boso} 
\begin{eqnarray}    
{\cal H}_{\rm ll} &\approx& V_{\rm ll} \int \frac{dx}{a_0}
\,\,C_{l1}\cos(\sqrt{8\pi}\phi_-)+\cdots,\label{4spin_boso_ll}   \\
{\cal H}_{\rm rr} &\approx& V_{\rm rr} \int \frac{dx}{a_0}\,\,
M^2 C_{r1}\cos(\sqrt{2\pi}\theta_-)\nonumber\\
&&+C_{r2}\cos(2\sqrt{2\pi}\theta_-)+C_{r3}\cos(\sqrt{8\pi}\phi_-)
\nonumber\\
&&+C_{r4}\cos(2\sqrt{8\pi}\phi_-)+\cdots,\label{4spin_boso_rr}\\
{\cal H}_{\times} &\approx& V_{\times} \int \frac{dx}{a_0}\,\,
-M^2 C_{\times 1}\cos(\sqrt{2\pi}\theta_-)\nonumber\\
&&+C_{\times 2}\cos(2\sqrt{2\pi}\theta_-)
+C_{\times 3}\cos(\sqrt{8\pi}\phi_-)
\nonumber\\
&&+C_{\times 4}\cos(2\sqrt{8\pi}\phi_-)+\cdots,\label{4spin_boso_dd}
\end{eqnarray}
\end{subequations}
where $C_{l1}$, $C_{rq}$ and $C_{\times q}$ are nonuniversal constants,
and we have written down only important terms. 
We note that (\rnum{1}) $C_{l1}$, $C_{rq}$, and $C_{\times q}$ with
$q=1,2$, and $4$ are all positive, 
(\rnum{2}) $C_{rq}$ is the same order as $C_{\times q}$ when $q=1,2$,
and $4$, and (\rnum{3}) $C_{r3}$ and $C_{\times 3}$ depend strongly on 
the magnetization $M$. In Eq.~(\ref{4spin_boso_ll}), we have used
the bosonized dimerization operator 
$(-1)^j{\bm S}_{n,j}\cdot{\bm S}_{n,j+1}
\sim \cos(\sqrt{4\pi}\phi_n+2\pi Mj)+\cdots$.~\cite{Gogo} 
The $\theta_+$ vertex operators and $\phi_+$ ones without oscillating
factors are absent in Eq.~(\ref{4spin_boso}) because the four-spin
exchanges~(\ref{4spin_terms}) conserve all the symmetries 
of the ladder~(\ref{S=1/2ladder}).
It is remarkable that ${\cal H}_{\rm rr}$ and ${\cal H}_{\times}$
include a new $\theta_-$ term, $\cos(2\sqrt{2\pi}\theta_-)$, which can
make the locking position of $\theta_-$ shift. From the calculation in
Eq.~(\ref{4spin_boso}), we find that a four-spin exchange 
term with the form 
\begin{eqnarray}
\label{cos_term}
({\bm S}_{1,i}\cdot{\bm S}_{2,j})({\bm S}_{1,k}\cdot{\bm S}_{2,l})
\end{eqnarray} 
always contains $\cos(2\sqrt{2\pi}\theta_-)$.~\cite{note_cos} 
The scaling dimension of $\cos(2\sqrt{2\pi}\theta_-)$ is
$2/K_-(\approx 2/K)$, and thus it becomes relevant when $K_->1$. 
Equation~(\ref{TLLpara_velocity}) and the property of $K$ suggest that 
the condition $K_->1$ is realized when the system is sufficiently close
to the saturated state ($K\to 1$) and the rung couping is AF ($J_\perp>0$).
If we introduce perturbations such as further-neighbor 
exchanges for the intrachain direction, 
the parameter regime with $K_->1$ could expand. 
In addition, four-spin exchanges would vary the value of $K_-$ slightly. 

Let us consider a spin ladder with 
four-spin terms $V_{\rm rr}$ and $V_\times$ under the condition
$K_->1$. In its bosonized Hamiltonian, the TLL of the 
$(\phi_+,\theta_+)$ sector is still stabilized by symmetries of the 
ladder. 
On the other hand, the $(\phi_-,\theta_-)$ sector is described as
\begin{subequations}
\label{Hami_4spin}
\begin{eqnarray}
\label{Hami_4spin_1}
{\cal H}_{[\phi_-,\theta_-]}\,\,=\,\,{\rm Gaussian\,\, part}
\hspace{3.5cm}\\
+v_-\int  dx \,\,
\frac{g_1}{\alpha^{d_1}}\cos(\sqrt{2\pi}\theta_-)
+\frac{g_2}{\alpha^{d_2}} \cos(2\sqrt{2\pi}\theta_-),
\nonumber\\
\left\{
\begin{array}{lll}
g_1&\propto& B_0^2J_\perp + M^2(C_{r1}V_{\rm rr}-C_{\times1}V_\times)\\
g_2&\propto&  C_{r2}V_{\rm rr}+C_{\times2}V_\times\\
\end{array}
\right.,\hspace{1cm}
\label{Hami_4spin_g} 
\end{eqnarray}
\end{subequations}
where $\alpha\sim a_0$ is the short-distance cut off, $g_{1,2}$ are
dimensionless coupling constants, $d_1=2-1/(2K_-)$ and $d_2=2-2/K_-$: 
$1/(2K_-)$ and $2/K_-$ are the scaling dimensions of 
$\cos(\sqrt{2\pi}\theta_-)$ and $\cos(2\sqrt{2\pi}\theta_-)$,
respectively. Here we have omitted irrelevant $\phi_-$ vertex terms. 
Equation~(\ref{Hami_4spin_1}) is a well-known 
double sine-Gordon (DSG) model.~\cite{dSG} 
It is widely believed that the DSG model exhibits an Ising-type
quantum phase transition. In fact, one easily finds that 
when $g_2$ is positively increased enough compared with $|g_1|$, 
the potential form of the DSG model changes from a single-well 
to a double-well type. The Ising transition takes
place just between the single-well (disordered) phase and 
the double-well (ordered) one. For the ordered phase, 
the locking position of $\theta_-$ deviates 
from $0$ and $\pm\sqrt{\pi/2}$. It indeed means the emergence 
of a finite vector chirality $\langle{\cal V}_j\rangle$ and 
the spontaneous breaking of the $Z_2$ rung-parity symmetry. 
Note that even in the chiral-ordered phase, the U(1) symmetry around the
$S^z$ axis ($\theta_+\to\theta_++$ const) is conserved. 
It is expected that even via the Ising transition, there is no singular
behavior in the magnetization curve since the DSG model in the
$(\phi_-,\theta_-)$ sector does not directly couple to the field $H$. 
Equation~(\ref{Hami_4spin_g}) indicates that the system favors 
the vector chiral long-range order, for example, under the
following conditions:
\begin{eqnarray}
\label{chirality_condition}
\begin{array}{lll}
(1)& V_{\rm rr}\agt J_\perp>0 \,\,\,{\rm and}\,\,\, V_\times\sim 0,\\
(2)& V_\times\agt J_\perp>0\,\,\, {\rm and}\,\,\, V_{\rm rr}\sim 0,\\ 
(3)& V_{\rm rr}\sim V_\times\agt J_\perp>0,
\end{array}
\end{eqnarray}
etc. It also tells us 
that the cyclic exchange term with $V_{\rm rr}=-V_\times$ 
possesses only a low possibility of causing the chiral order. 
From these arguments based on the bosonization, 
we conclude that the vector chiral long-range order emerges 
in the field-induced TLL, when a sufficiently strong 
four-spin exchange~(\ref{cos_term}) is introduced in the ladder 
with $K_->1$. 
Furthermore, if $K_->1$ and $M=0$ ($H<H_l$) simultaneously hold in a
U(1)-symmetric ladder, 
i.e., if $K_->1$ holds when the $(\phi_+,\theta_+)$ sector 
has a gapped spectrum (although this situation is hard to occur 
within a realistic ladder system), 
it is also possible that a four-spin exchange (\ref{cos_term}) 
yields a {\it gapped} chiral-ordered state.

\begin{figure}[t]
\scalebox{0.7}{\includegraphics[width=\linewidth]{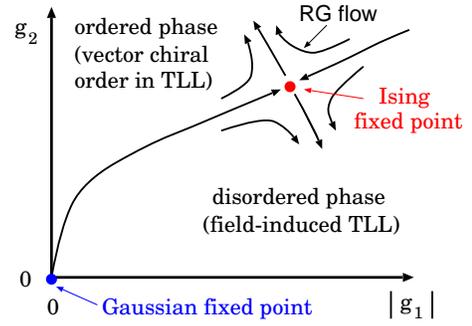}}
\caption{\label{RGflow_dSG} RG flow and phase diagram of the DSG model 
with $K_->1$ around the Gaussian fixed point $(g_1,g_2)=(0,0)$. 
Phrases in the parentheses stand for the phases of the corresponding 
spin-$\frac{1}{2}$ or spin-1 AF ladders with four-spin exchanges and 
a Zeeman term: $g_{1,2}$ are given by Eq.~(\ref{Hami_4spin_g}) 
[Eq.~(\ref{g_S=1})] in the spin-$\frac{1}{2}$ [spin-1] case. 
Near the Gaussian point, the RG-invariant curve satisfies 
$g_2\sim |g_1|^{d_2/d_1}$. 
The RG analysis of the DSG model has been discussed in 
Refs.~\onlinecite{Tsuchi} and \onlinecite{Ye}.}
\end{figure}

In order to see the global phase structure of the DSG model 
around the trivial Gaussian fixed point $(g_1,g_2)=(0,0)$, 
the renormalization-group (RG) analysis is very useful. 
Using the operator product expansion technique,~\cite{Cardy} we
obtain the following one-loop RG equation for 
the DSG model:~\cite{Tsuchi,Ye}  
\begin{eqnarray}
\label{RG_dSG}
\frac{dg_1}{dL}& = & \Big(2-\frac{1}{2K_-}\Big)g_1-\pi g_1g_2,
\nonumber\\
\frac{dg_2}{dL}& = & \Big(2-\frac{2}{K_-}\Big)g_2-\pi g_1^2,
\end{eqnarray}
where $L$ is the scaling parameter: $\alpha\to\alpha e^L$. 
The RG equations have two nontrivial fixed points 
$(\pm\sqrt{d_1 d_2}/\pi,d_1/\pi)=(\pm g_1^*,g_2^*)$. 
These must correspond to the above-mentioned Ising transition fixed
point although their values $g_{1,2}^*$ would not be reliable. 
Analyzing the RG equations around $(g_1,g_2)=(0,0)$ and $(\pm g_1^*,g_2^*)$,
we can draw the RG flow and the ground-state phase diagram as in
Fig.~\ref{RGflow_dSG}. As expected, it shows that if $g_2$ sufficiently
grows, the vector chiral order appears.

The vector chirality ${\cal V}_j$ must play the role of the order
parameter near the Ising transition. Therefore applying known results 
of the two-dimensional Ising model~\cite{WuMc,Shel} we can predict
several properties of the chirality ${\cal V}_j$.  
For instance, the chirality increases as 
$\langle{\cal V}_j\rangle\sim(g_2-g_2^c)^{1/8}$ 
in the vicinity of the critical point $(g_1^c,g_2^c)$.
In addition, the chirality correlation function near the critical point 
is predicted to behave as follows: 
\begin{eqnarray}
\label{chiral_corre}
\lim_{j\to\infty}\langle{\cal V}_j{\cal V}_0\rangle \approx
\hspace{5.5cm}\nonumber\\
\left\{ 
\begin{array}{lcc}
C_1 e^{-|j|/\xi_1}/\sqrt{|j|}  & (g_2<g_2^c) \\
C_0/|j|^{1/4}  &  (g_2=g_2^c)\\
C_3(g_2-g_2^c)^{1/4}+C_2 e^{-|j|/\xi_2}/\sqrt{|j|} & (g_2>g_2^c) 
\end{array}
\right., 
\end{eqnarray}
where $C_n$ are nonuniversal constants, and $\xi_{1(2)}$ is the 
correlation length in the disordered (ordered) phase of the DSG model. 
The result~(\ref{chiral_corre}) would be useful in numerically
detecting the chiral-ordered phase. 
Furthermore, if the chiral-ordered phase is realized in a ladder compound,
these features of the vector chirality could be observed experimentally, 
in principle.

As we already mentioned in the Introduction, in spin-$\frac{1}{2}$
ladder systems, there is a duality relation that has been studied 
in Ref.~\onlinecite{Hi-Mo}. 
Let us here apply the duality to our spin-$\frac{1}{2}$ ladders 
with four-spin exchanges. The duality transformation is defined as 
${\bm S}_{1,j}=\frac{1}{2}({\bm T}_{1,j}+{\bm T}_{2,j})
+{\bm T}_{1,j}\times {\bm T}_{2,j}$ and 
${\bm S}_{2,j}=\frac{1}{2}({\bm T}_{1,j}+{\bm T}_{2,j})
-{\bm T}_{1,j}\times {\bm T}_{2,j}$. New operators ${\bm T}_{n,j}$
obey the same algebra as spin-$\frac{1}{2}$ operators. 
This mapping therefore makes an arbitrary spin-$\frac{1}{2}$ ladder 
with $\{{\bm S}_{n,j}\}$ change into a new
dual ladder with $\{{\bm T}_{n,j}\}$. 
It is remarkable that the duality transformation 
conserves the total spin in each rung 
($\bm{S}_{1,j}+\bm{S}_{2,j}=\bm{T}_{1,j}+\bm{T}_{2,j}$) 
and changes the vector chirality in each rung to a N\'eel-type moment 
($-2\bm{S}_{1,j}\times \bm{S}_{2,j}=\bm{T}_{1,j}-\bm{T}_{2,j}$). 
On the other hand, recall that a spin-$\frac{1}{2}$ ladder 
${\cal H}_*={\cal H}_{\rm lad}[J,J_\perp]
+{\cal H}_{\rm rr}[V_{\rm rr}]+{\cal H}_{\times}[V_\times]$
is shown to possess the vector chiral long-range order 
near the saturation if the condition $K_->1$ is satisfied. 
Combining this prediction and the above
properties of the duality, we can find that 
the dual model for ${\cal H}_*$ has a rung N\'eel order 
$\langle T_{1,j}^z-T_{2,j}^z\rangle\neq 0$ in the vicinity of the
saturation ($T_{1,j}^z+T_{2,j}^z\to 1$). The Hamiltonian of the dual
ladder is represented as 
\begin{eqnarray}
\label{dual_model}
{\cal H}_{\rm dual}&=&
(\frac{J}{2}-\frac{V_{\times}}{8})
\sum_{n,j}\bm{T}_{n,j}\cdot\bm{T}_{n,j+1} 
+J_\perp\sum_{j}\bm{T}_{1,j}\cdot\bm{T}_{2,j}\nonumber\\
&&
+(\frac{J}{2}+\frac{V_{\times}}{8})
\sum_{j}\bm{T}_{1,j}\cdot\bm{T}_{2,j+1}
+\bm{T}_{2,j}\cdot\bm{T}_{1,j+1}\nonumber\\
&&
+V_{\rm rr}\sum_{j}(\bm{T}_{1,j}\cdot\bm{T}_{2,j})
(\bm{T}_{1,j+1}\cdot\bm{T}_{2,j+1}) \nonumber\\
&&
+(2J+\frac{V_\times}{2})\sum_{j}(\bm{T}_{1,j}\cdot\bm{T}_{1,j+1})
(\bm{T}_{2,j}\cdot\bm{T}_{2,j+1}) \nonumber\\
&&
+(-2J+\frac{V_\times}{2})\sum_{j}(\bm{T}_{1,j}\cdot\bm{T}_{2,j+1})
(\bm{T}_{2,j}\cdot\bm{T}_{1,j+1}) \nonumber\\
&&
-H\sum_{n,j}T_{n,j}^z.
\end{eqnarray}
Since this model has a strong rung-coupling term and strong 
four-spin ones, it cannot be analyzed within the weak-rung-coupling
approach in this paper.

\section{Spin-$\bm 1$ ladders} 
\label{4spin_spin1} 
Let us now turn to the two-leg spin-1 ladder~(\ref{S=1/2ladder}) 
where spin-$\frac{1}{2}$ operators are replaced with spin-1 ones. 
One will encounter a scenario similar to the 
spin-$\frac{1}{2}$ case below.

In the decoupled case with zero field,
each AF chain has a Haldane gap, $\simeq 0.41J$. 
When the magnetic field $H$ exceeds it, a field-induced TLL
appears due to the $S^z=1$ magnon condensation. 
For this TLL phase, the effective field theory has been
established.~\cite{S1eff,S1eff_NLSM,S1eff_EA,MS06} 
The effective Hamiltonian consists of the Gaussian part,
which corresponds to the TLL, plus the part of $S^z=0$ and $-1$ massive 
magnons. In contrast to the spin-$\frac{1}{2}$ case, 
the TLL parameter $K$ is larger than 1.~\cite{Fath} 
The effective theory enables us to write the field theory form of
spin operators as follows:
\begin{subequations}
\label{boso_spin_1} 
\begin{eqnarray}
S^z_{n,j} &\sim&
M-a_0\partial_x\phi_n/\sqrt{\pi}+D_1\cos(\sqrt{4\pi}\phi_n-2\pi Mj)
\nonumber\\
&&+(-1)^j[D_2 \sigma_n\cos(\sqrt{\pi}\phi_n-\pi Mj)
\nonumber\\ 
&&+D_3(\eta_n e^{i\sqrt{\pi}\theta_n}+{\rm h.c.})]+\cdots,
\end{eqnarray}
\begin{eqnarray}
S^+_{n,j}&\sim& (-1)^j e^{-i\sqrt{\pi}\theta_n}\mu_n
[E_2+E_3\cos(\sqrt{4\pi}\phi_n-2\pi Mj)]\nonumber\\
&&+E_1 e^{-i\sqrt{\pi}\theta_n}
\cos(\sqrt{\pi}\phi_n-\pi Mj)(\xi_{L,n}+i\xi_{R,n})
\nonumber\\
&&+\cdots,
\end{eqnarray}
\end{subequations} 
where $D_q$ and $E_q$ are nonuniversal constants, $(\phi_n,\theta_n)$
are the massless boson fields for the $S^z=1$ condensed magnon, 
the set $(\xi_{\nu,n},\sigma_n,\mu_n)$ describes the $S^z=0$ magnon 
sector, and $\eta_n$ is the $S^z=-1$ magnon field. 
For the massive $S^z=0$ magnon sector, 
$\langle \sigma_n\rangle =0$ and $\langle\mu_n\rangle\neq 0$ hold. 
[For a more detailed explanation of Eq.~(\ref{boso_spin_1}), 
see Ref.~\onlinecite{MS06}.] If we consider only 
part of the TLL sector, the symmetries of the spin-1 ladder can be 
expressed by fields $\phi_n$ and $\theta_n$: a U(1) rotation
around the $S^z$ axis and the one-site translation
respectively correspond to 
$\theta_n\to \theta_n+{\rm const}$ and 
$(\phi_n,\theta_n)\to(\phi_n-M\sqrt{\pi},\theta_n+\sqrt{\pi})$.~\cite{MS06}
These symmetry operations are very similar to 
those of the spin-$\frac{1}{2}$ case.

Substituting the formula~(\ref{boso_spin_1}) into the rung
coupling of the spin-1 ladder~(\ref{S=1/2ladder}) and 
integrating out the massive-magnon part in its effective
Hamiltonian,~\cite{note} we obtain a two-component boson 
theory that is the same type as Eq.~(\ref{S=1/2eff}). 
Namely, the $(\phi_+,\theta_+)$ sector again
brings a TLL, and the $(\phi_-,\theta_-)$ sector takes a massive
spectrum with pinning $\theta_-$ to $0$ or $\pm\sqrt{\pi/2}$.~\cite{MS05} 
The TLL state is strongly protected by symmetries of the ladder like
the case of the spin-$\frac{1}{2}$ ladder. 
Formula (\ref{boso_spin_1}) 
further tells us that the vector chirality is written as 
\begin{eqnarray}
\label{chiral_S1}
{\cal V}_j\sim \mu_1\mu_2\sin(\sqrt{2\pi}\theta_-)+\cdots,
\end{eqnarray} 
and leading parts of ${\cal N}_{\pm,j}$ are written 
as a function of $\phi_{\pm}$.
These field theory results are indeed analogous to those 
of the spin-$\frac{1}{2}$ case, Eq.~(\ref{orderpara_boso}).  
The formula~(\ref{chiral_S1}) shows that $\langle {\cal V}_j\rangle=0$
holds in the standard spin-1 ladder~(\ref{S=1/2ladder}) with $M\neq 0$.

For the TLL phase of the spin-1 ladder, 
let us add a four-spin interaction term. 
In this paper, we focus on a so-called biquadratic term~\cite{M-K,Lu-Sch} 
\begin{eqnarray}
\label{biquadratic}
{\cal H}_b &=& V\sum_j({\bm S}_{1.j}\cdot{\bm S}_{2,j})^2,
\end{eqnarray}
that is one of realistic, familiar four-spin terms in spin-1 systems.  
Employing the field theory formula~(\ref{boso_spin_1}) and 
symmetry arguments, and then tracing out massive magnons, 
we obtain the bosonized form of ${\cal H}_b$, which contains 
$\cos(2\sqrt{2\pi}\theta_-)$. As a result, 
the $(\phi_-,\theta_-)$ sector including effects of 
${\cal H}_b$ could be described by a DSG model that has the same form
as Eq.~(\ref{Hami_4spin_1}). For the present case, the coupling constants
$g_{1,2}$ are evaluated as 
\begin{eqnarray}
\label{g_S=1}
g_1 &\propto &  \tilde{C}_1 J_\perp +M^2 \tilde{C}_2 V+\cdots, 
\nonumber\\
g_2 &\propto& \tilde{C_3}V +{\cal O}_1\left(\frac{V}{J}\right)V
+{\cal O}_2\left(\frac{J_\perp}{J}\right)J_\perp,
\end{eqnarray}
where $\tilde{C}_q$ are nonuniversal positive constants, 
and the second and third terms of $g_2$ 
originate from the trace-out procedure of massive magnons.
As we already mentioned, since the TLL parameter $K$ of the spin-1 AF chain
is larger than 1, $K_->1$ would hold in a wide parameter regime where 
$\cos(2\sqrt{2\pi}\theta_-)$ is relevant. 
From Eq.~(\ref{g_S=1}) and Fig.~\ref{RGflow_dSG}, we can conclude that 
if $g_2\agt |g_1|$, namely, $V\agt |J_\perp|$, 
under the condition $K_->1$, 
a vector chiral long-range order emerges like the spin-$\frac{1}{2}$
case. In addition, Eq.~(\ref{chiral_S1}) indicates that around the Ising
transition, the chirality correlation function in spin-1 ladders 
behaves as Eq.~(\ref{chiral_corre}). 
As one easily expects from Eqs.~(\ref{cos_term}) and (\ref{boso_spin_1}), 
besides the biquadratic term, 
other four-spin exchanges with the form~(\ref{cos_term}) 
can also cause the vector chiral order.

\section{Parity-breaking perturbations}
\label{chiral_break}
From Sec.~\ref{bos_review} to Sec.~\ref{4spin_spin1}, 
we have studied two-leg spin ladders with the $Z_2$
rung-parity symmetry. In this section, we shortly discuss effects of 
rung-parity-breaking perturbations. Let us focus on the following 
three kinds of realistic perturbations for spin-$\frac{1}{2}$ ladders:
\begin{subequations}
\label{parity_break}
\begin{eqnarray}
\label{break1_DM}
{\cal H}_{\rm uDM} &=& \sum_j D_u({\bm S}_{1,j}\times{\bm S}_{2,j})^z,
\\
\label{break2_DM}
{\cal H}_{\rm sDM} &=& \sum_j (-1)^jD_s({\bm S}_{1,j}\times{\bm S}_{2,j})^z,
\\
\label{break3_threespin}
{\cal H}_{\rm chiral} &=& \sum_{p,q,r}A_{pqr}
{\bm S}_{p}\cdot({\bm S}_{q}\times{\bm S}_{r}),
\end{eqnarray}
\end{subequations}
where $p$, $q$, and $r$ in Eq.~(\ref{break3_threespin}) represent 
a site index on a spin ladder, $\{n,j\}$.
The uniform DM term~(\ref{break1_DM}) and the staggered
one~(\ref{break2_DM})~\cite{DMterm} often emerge 
in sufficiently low-crystal-symmetric magnets, and their origin is 
spin-orbit coupling. 
The third three-spin 
term~(\ref{break3_threespin})~\cite{three_spin,three_spin_2} might be
unfamiliar. It originates from virtual electron-hopping processes on
a triangle plaquette $\{p,q,r\}$ in a Mott-insulating ladder 
under an external magnetic field. The coupling constant 
$A_{pqr}\propto \sin(e\Phi_{pqr})$, where 
$e$ is the electron charge and $\Phi_{pqr}$ is the magnetic flux 
enclosed by the plaquette $\{p,q,r\}$. 
It therefore depends heavily on the direction and
the strength of the magnetic field, and vanishes when the plaquette plane
and the direction of the magnetic field is parallel or the magnetic
field is absent. (For a more detailed discussion on the three-spin term,
see Refs.~\onlinecite{three_spin} and \onlinecite{three_spin_2}.)

Formula~(\ref{boso_spin_1/2}) allows us to bosonize the DM terms 
as follows: 
\begin{subequations}
\label{break_DM}
\begin{eqnarray}
\label{break_uDM}
{\cal H}_{\rm uDM} &\approx& -D_u \int \frac{dx}{a_0} B_0^2
\sin(\sqrt{2\pi}\theta_-)+\cdots,\\
\label{break_sDM}
{\cal H}_{\rm sDM} &\approx& -D_s \int \frac{dx}{a_0} (-1)^j B_0^2
\sin(\sqrt{2\pi}\theta_-)+\cdots.\hspace{0.6cm}
\end{eqnarray}
\end{subequations}
This result is obvious from the bosonized expression of vector
chirality, Eq.~(\ref{orderpara_boso_1}). As we already discussed, the
phase field $\theta_-$ is locked at $0$ or $\pm\sqrt{\pi/2}$ in the 
spin ladder ${\cal H}_{\rm lad}$ due to the potential 
$\cos(\sqrt{2\pi}\theta_-)$. Equation~(\ref{break_uDM}) clearly shows 
that as soon as a uniform DM term is added to ${\cal H}_{\rm lad}$, 
the locking position of $\theta_-$ immediately
varies and the vector chirality takes a finite expectation value.   
In contrast, since a small staggered DM term is irrelevant 
due to the factor $(-1)^j$, it hardly affects the low-energy physics
of the ladder ${\cal H}_{\rm lad}$.

Similarly, one can bosonize 
the three-spin term~(\ref{break3_threespin}). 
It is readily found that Eq.~(\ref{break3_threespin}) also contains 
$\sin(\sqrt{2\pi}\theta_-)$, and is hence expected to cause a 
finite vector chirality. However, one should note that 
besides the sine term, the three-spin term includes 
other relevant or marginal terms such as 
$\partial_x\phi_\pm\partial_x\theta_\mp$. Therefore 
effects of the three-spin term would not be simply argued. 
For instance, authors in Ref.~\onlinecite{three_spin_2} have obtained
an unexpected prediction that for spin-$\frac{1}{2}$ easy-plane zigzag
ladder, whose ground state has a vector chiral order, 
a three-spin term helps the chiral order vanish. 
The full bosonized form of Eq.~(\ref{break3_threespin}) and its effects
on spin ladders depends on the coupling constant 
$A_{pqr}$, namely, the underlying Mott-insulating ladder. 
In spin-1 or higher-spin ladder systems, parity-breaking terms such
as Eq.~(\ref{parity_break}) would be also able to appear, and they
induce a finite vector chirality.

From the arguments in this section, we can expect that
if a small parity-breaking term like Eq.~(\ref{break1_DM}) is present 
in spin ladders with four-spin exchanges, 
it could be a trigger of the emergence of the
four-spin-exchange- and magnetic-field-driven vector chiral order. 
Furthermore, such a term 
would help the chiral order survive even at a finite temperature.

\section{Conclusions and Discussions}
\label{final}
We have studied effects of four-spin exchanges in
two-leg spin-$\frac{1}{2}$ and spin-1 AF ladders under a magnetic field, 
utilizing bosonization techniques. The magnetic field reduces the spin
SU(2) symmetry to the U(1) one, and as a result, the emergence of vector
chirality $\langle {\cal V}_j\rangle$ becomes possible. 
In Secs.~\ref{4spin_spin1_2} and \ref{4spin_spin1}, 
certain types of four-spin exchanges with the form~(\ref{cos_term}) are 
indeed shown to yield the vector chiral long-range order 
$\langle{\cal V}_j\rangle\neq 0$ 
with the $Z_2$ rung-parity symmetry spontaneously broken and 
the U(1) one unbroken, 
if the four-spin exchanges are large enough and 
the TLL parameter $K_-$ is larger than $1$. 
Spin-1 ladders more easily satisfy the condition $K_->1$
rather than spin-$\frac{1}{2}$ ones: for the spin-$\frac{1}{2}$ case, 
$K_->1$ is predicted to hold only near the saturated state, 
while for the spin-1 case, it would hold in a wide region 
from the lower critical magnetic field to the upper one. 
An Ising-type phase transition takes place between the 
vector-chiral-ordered state and the standard spin liquid. 
Around the transition, the vector chirality correlation function follows
Eq.~(\ref{chiral_corre}). The Ising transition 
would not accompany any singularity of the magnetization
curve, because the $(\phi_-,\theta_-)$ sector, which brings the
transition, does not directly couple to the magnetic field $H$.
It is noteworthy that for spin-$\frac{1}{2}$ ladders with 
a four-spin {\it cyclic} exchange, the chiral order is hard to
generate. For spin-$\frac{1}{2}$ ladders with four-spin exchanges, 
we have also applied the duality transformation.~\cite{Hi-Mo} The dual spin
ladder~(\ref{dual_model}) is predicted to possess a N\'eel order for the
rung direction. In Sec.~\ref{chiral_break}, we have briefly considered 
effects of rung-parity-breaking perturbations on spin ladders. 
In particular, we clearly find that 
a uniform DM interaction~(\ref{break1_DM}) yields a
finite expectation value of vector chirality in the ladder 
${\cal H}_{\rm lad}$ even if it is small, while a staggered DM 
term~(\ref{break2_DM}) is irrelevant in the same ladder. 
Moreover, it is expected that some types of these perturbations
including Eq.~(\ref{break1_DM}) could play the role of the trigger 
of the vector chiral order generated by four-spin exchanges 
and a magnetic field.

Our predictions in the spin-$\frac{1}{2}$ and spin-1 cases suggest that 
for general spin-$S$ AF ladders, a four-spin exchange such as 
Eq.~(\ref{cos_term}) and a magnetic field can also induce a vector chiral
order. Within the bosonization framework, it is difficult to
quantitatively calculate critical values of the four-spin exchanges: 
$V_{\rm rr}^c$, $V_{\times}^c$, and $V^c$. Determining them accurately
is an interesting future problem.

Unfortunately, the predicted vector chiral long-range order 
has never been detected in experiments. 
However, since it is suggested~\cite{M-K} that 
effects of four-spin exchange cannot be negligible in some magnets 
and a magnetic field is one of the few things that we can control, 
there is a possibility that an experimental 
evidence of the chiral order is found. 
It is shown that a polarized neutron scattering experiment 
can detect a vector chiral order.~\cite{Mal} 
In addition, if four-point spin correlation functions such as 
nonlinear susceptibilities are experimentally 
observed, 
they must contain some 
features of the vector chiral order or correlation 
[see Eq.~(\ref{chiral_corre})].

As is well known, theoretically, a typical origin of four-spin interactions 
is fourth- or higher-order electron-hopping processes in Mott
insulators.~\cite{Taka-Yoshi} Hence the low-energy physics 
of Mott insulators with a large hopping matrix element 
would be described by a spin system with 
considerably large four-spin terms. Recently 
techniques in optical lattice have been greatly developed and 
they make it possible to construct several insulating states.~\cite{Ripoll} 
Experiments in optical lattice therefore might succeed in realizing 
a vector-chiral-ordered state. 
It is also known that spin-phonon couplings can induce four-spin
terms.~\cite{spin-phonon} Therefore magnets with such a strong coupling 
might be another candidate for a chiral-ordered state.

It is shown in a previous work~\cite{Tots} 
that in spin-$\frac{1}{2}$ ladders, 
a four-spin cyclic exchange can yield a 
$M=1/4$ plateau state with a N\'eel-type order 
when $J_\perp\gg J$. In our weak-rung-coupling approach, 
such a plateau phase could also appear if a vertex operator 
$\cos(2\sqrt{8\pi}\phi_+ +8\pi Mj)=\cos(2\sqrt{8\pi}\phi_+)$ 
in the $(\phi_+,\theta_+)$ sector becomes
relevant, i.e., $K_+$ is smaller than $1/4$. Supposing that 
$K_+<1/4$ and $K_->1$ simultaneously hold 
in the weak-rung-coupling regime, 
we can expect that the vector chiral order, caused by 
four-spin exchanges, still survives in the plateau state. 
In this plateau regime, there is the possibility that 
a subleading term of ${\cal V}_j$, 
$\sin(\sqrt{2\pi}\theta_-)\cos(\sqrt{8\pi}\theta_+ +4\pi Mj)$, 
causes a staggered vector chiral order. 
We, however, note that the conditions $K_+<1/4$ and $K_->1$ are 
difficult to occur in real ladders.

Spin-$\frac{1}{2}$ ladders with four-spin exchanges possess 
an exactly solvable SU(4)-symmetric point, 
around which we can construct its effective field theory.~\cite{Aza,Le-To} 
The relationship between the SU(4)-point field theory and the bosonized
one in the present paper has not been established well. 
Making it clear is an interesting open problem, and must contribute to a 
more sophisticated understanding of spin-$\frac{1}{2}$ ladders. 

We finally note that besides a magnetic field, other U(1)-symmetric
terms such as $XXZ$ anisotropy would help the emergence of the chiral
order in AF spin ladders with four-spin exchanges.

\begin{acknowledgments}
The author thanks Masahisa Tsuchiizu for useful comments on 
the DSG model. This work was supported by a Grant-in-Aid for Scientific 
Research (B) (No. 17340100) from the Ministry of Education, 
Culture, Sports, Science and Technology of Japan.
\end{acknowledgments}


\end{document}